\documentclass[12pt]{article}
\usepackage{amssymb}
\usepackage{setspace}
\newcommand{\bra}[1]{\langle #1|}
\newcommand{\ket}[1]{|#1\rangle}
\newcommand{\braket}[2]{\langle #1|#2\rangle}
\newcommand{\be}{\begin{eqnarray}}
\newcommand{\ee}{\end{eqnarray}}

\newcommand{\co}[2]{\lbrack #1 , #2 \rbrack}

\newcommand{\p}{\partial}

\newcommand{\id}{\mathbb{I}}
\newcommand{\mbold}[1]{\mbox{\boldmath ${#1}$}}

\begin{document}


\def\maketitle{
   \begin{flushright}
   hep-th/0507242\\  DAMTP-2005-68
   \end{flushright}
   \begin{center}
   \Huge{Operator Algebra in Logarithmic Conformal Field Theory}\\
   \vskip 0.7 cm
   \large{Jasbir Nagi}\\ \vskip 0.3 cm \large{J.S.Nagi@damtp.cam.ac.uk}\\ \vskip 0.3 cm \large{DAMTP,
   University of Cambridge, Wilberforce Road,}\\ \vskip 0.3 cm \large{Cambridge,
    UK, CB3 0WA}
   \vskip 0.7 cm
   \end{center}
   }

\maketitle
\begin{abstract}
For some time now, conformal field theories in two dimensions have
been studied as integrable systems. Much of the success of these
studies is related to the existence of an operator algebra of the
theory. In this note, some of the extensions of this machinery to the
logarithmic case are studied, and used. More precisely, from M\"obius symmetry
constraints, the generic three and four point functions of logarithmic
quasiprimary fields are calculated in closed form for arbitrary Jordan
rank. As an example, $c=0$ disordered systems with non-degenerate vacua are
studied. With the aid of two, three and four point functions, the
operator algebra is obtained and associativity of the algebra studied.
\end{abstract}

\onehalfspacing

\newpage
\section{Introduction}

Since the studies of Belavin, Polyakov and Zamolodchikov \cite{bpz} and
Zamolodchikov and Zamolodchikov \cite{zz}, Conformal Field Theory in
two dimensions (2dCFT) has been heavily studied, in particular as an
integrable quantum field theory. This means that, in principle, the
model is exactly solvable, i.e. all of the correlation functions can
be determined, and hence the S-matrix. Central to this idea is the
notion of an `operator algebra', where the operators of the theory are
endowed with a product, so that a product of two operators may be
realized as a linear combination of other operators in the
theory. Thus, the calculation of an $N$-point function can be reduced
to one of a linear combination of  $N-1$ point functions. Given a quantum field theory, one
might ask - what is the operator algebra? In general, this is a
difficult question to answer, but in 2dCFT, \cite{bpz} demonstrated
how to obtain it. The method depends crucially on the form of two,
three and four point functions of `primary' fields of the
theory. Primary fields of the theory are fundamental fields of the
theory, in the sense that all of the other fields can be obtained by
action of the symmetry algebra on the primary fields.

Logarithmic conformal field theories (LCFTs) are a class of 2dCFT that have
been studied heavily over the last 10 years. They differ from
more traditional 2dCFTs by having logarithms in correlation functions,
and having elements of the Cartan subalgebra which are non-diagonalizable - most
notably the dilation operator $L_0$ is non-diagonalizable, and hence
must be represented in terms of Jordan blocks. Due to these
complexities, obtaining an understanding of the $2,3,4$ point functions and the
operator algebra has been tricky, although much work has been done
towards this, including,
\cite{flohr1}\cite{flo3}\cite{flo}\cite{flohr2}\cite{flohr3}\cite{kogannich1}\cite{kogannich2}\cite{moghi}\cite{nagi1}.

In this note, using the constructions developed in \cite{nagi1} (which
are reviewed in section 2), the
three and four point functions of logarithmic quasiprimary fields are calculated (sections 3 and 4 respectively). It should be stressed
that in the calculation of these correlators, no assumptions on the
operator algebra are made, and no ansatz is used, all of which are
often used in the literature. The calculations are
valid for arbitrary Jordan block size, even when the different Jordan
blocks inside the correlator are of different rank. The only input is
the conformal Ward identities corresponding to the M{\"o}bius group,
which are subsequently solved for, yielding general solutions.

In order to try and find an operator algebra using these $2,3,4$ point
functions, the case of $c=0$ systems with non-degenerate vacua (that
is, the vacuum does not have a logarithmic partner - for the degenerate
vacua case, see \cite{flohr4}\cite{kogannich2}) is analyzed, which pertains to
two dimensional systems with quenched disorder
\cite{ludgur2}\cite{ludgur}. This is an unusual LCFT, in that the
vacuum does not belong to a Jordan block, whereas the stress energy
tensor does. Indeed, there are questions as to whether or not such an
unusual LCFT can be consistent. In section 5, it is found that from just
assuming that the vacuum is non-degenerate, that $c=0$, and that the
stress energy tensor (a.k.a. the Virasoro generator) has a
logarithmic partner (\ref{TT}), (\ref{Tt}), with help from the three
point function of section 3, one is, remarkably, able to find the entire operator
algebra, which corresponds to the one found in \cite{kogannich2}. In
section 6, a partial study of associativity is then conducted, which, as is the
norm in 2dCFT, comes down to studying four point functions. No
inconsistencies are found.

\section{Review of Logarithmic Primaries}

Logarithmic conformal field theories are conformal field theories that
are characterized by $L_0$ being non-diagonalizable and logarithms
appearing in correlation functions. To this end, one can try and alter
the definition of a primary field to accommodate the non-diagonal
behaviour, and see if logarithms come out. The author should stress that
it is not obvious to him if all logarithmic conformal field
theories can be realized in this way.

Consider an action of the Virasoro algebra on fields $\phi_i(z)$,
$i=0\ldots N-1$ given by $m\in\mathbb{Z}$,
\be\label{naiveprim}
\co{L_m}{\phi_i(z)} &=& z^m(m+1)(h\phi_i(z) + \phi_{i+1}(z)) +
z^{m+1}\p\phi_i(z)\qquad i=0\ldots N-2\\ \label{naiveprim2}
\co{L_m}{\phi_{N-1}(z)} &=& z^m(m+1)h\phi_{N-1}(z) +
z^{m+1}\p\phi_{N-1}(z).
\ee
Now, the $\phi_{i+1}$ term in (\ref{naiveprim}) prevents this from
being a collection of $N$ primary fields of conformal weight $h$ - indeed as it stands there
is only one primary field. Acting on the vacuum $\ket{0}$, and
considering $z=0$, one finds
\be
L_0\ket{\phi_i} = h\ket{\phi_i} + \ket{\phi_{i+1}}\textrm{  for }i=0\ldots
N-2, \qquad L_0\ket{\phi_{N-1}} = h\ket{\phi_{N-1}}
\ee
and thus the primary field corresponds to the eigenvector of the
Jordan block. In light of this, one can construct a vector $v_\phi (z)$ out of the
$\phi_i(z)$, and rewrite (\ref{naiveprim})(\ref{naiveprim2}) as
\be\label{vectorprim}
\co{L_m}{v(z)} = z^m(m+1)(h+J)v_\phi (z) + z^{m+1}\p v_\phi (z)
\ee
where $J$ is a rank $N$ nilpotent matrix, that is satisfies $J^N=0$,
$J^{N-1}\neq 0$. Now, one might try to integrate up
(\ref{vectorprim}), to obtain a geometric object, and one finds that
$v(z)$ can be realized as a section of the formal rank $n$
vector bundle whose transition functions are generated by $dz^{h+J}$
(see \cite{nagi1} for more details). Now, given transition functions
for a vector bundle, one is always free to rewrite everything in terms
of $G$-bundles. In the case at hand, this translates to defining
\be
\phi (z,J) := \sum_{i=0}^{N-1}\phi_i(z) J^{N-i-1}
\ee 
in which case (\ref{vectorprim}) reads
\be\label{Jprim}
\co{L_m}{\phi (z,J)} = z^m(m+1)(h+J)\phi (z,J) + z^{m+1}\p\phi (z,J)
\ee
and defines a logarithmic primary field $\phi$ of weight $h$ and rank $N$. If (\ref{Jprim}) only holds for $m\in\{ -1,0,1\}$, then $\phi$ is a logarithmic quasiprimary field.
It should be emphasized that (\ref{Jprim}), (\ref{vectorprim}) and the
pair (\ref{naiveprim})(\ref{naiveprim2}) are equivalent ways of
describing the same thing. Whilst (\ref{vectorprim}) might be a more convenient
realization when studying representation theory, (\ref{Jprim}) is more
convenient for studying the operator algebra and correlation
functions, and hence will be used here. 

As is usual in conformal field theory, one can restrict to $m\in\{
1,0,-1\}$ to obtain the action under the Lie algebra of the M{\"o}bius
group. Since the $L_0, L_{\pm 1}$ annihilate the vacuum, these can be
used to give readily solvable Ward identities for the correlation
functions. For example, one has on the two point function $\langle
\phi (z,J)\otimes\psi (w,K)\rangle$,
\be
\langle\co{L_m}{\phi (z,J)}\otimes\psi (w,K)\rangle + \langle\phi
  (z,J)\otimes\co{L_m}{\psi (w,K)}\rangle =0
\ee
which can be solved \cite{nagi1} to yield
\be\label{gen2ptfn}
 \langle\phi (z,J)\otimes\psi (w,K)\rangle = \mbold{C}(J,K)(z-w)^{-2(\id\otimes\id h_1 + J\otimes\id )}
\ee
where, for the two point function to be non-zero, one must have the
conformal weights of $\phi$ and $\psi$ identical, i.e. $h_1-h_2=0$, as well as
$(J-K)\mbold{C}=0$, where $\mbold{C}$ is a `function' of the $J,K$,
i.e. has an expansion
\be
\mbold{C} = \sum_{i=0, j=0}^{N-1, M-1}C_{i,j}J^i\otimes K^j.
\ee
Note, for the particular case of $N=M=2$, that is $J^2=0=K^2$, one has
the two point function (surpressing tensor products)
\be\label{2ptfn}
\langle\phi (z,J)\psi (w,K)\rangle = (z-w)^{-2h_1}\Big( (J+K)a + JK(b-2a\log (z-w))\Big)
\ee
where $a$ and $b$ are arbitrary, yielding the logarithms, as
promised. For higher rank Jordan blocks, the solution, when expressed
in components, can become quite unwieldy, with powers of logarithms all over the
place. For the remainder of this note, the tensor products will be surpressed for clarity.

\section{Three point function}
Consider the three point function
\be
\langle \phi_1(x,J)\phi_2(y,K)\phi_3(z,L)\rangle = \mbold{f}(x,y,z,J,K,L)
\ee
where $J^M=0$, $J^{M-1}\neq 0$, $K^N=0$, $K^{N-1}\neq 0$, $L^P=0$ and
$L^{P-1}\neq 0$ for some $M,N,P\in\mathbb{Z}$ with $M,N,P\geq 2$. For the purposes of this calculation, the
co-ordinates
\be
t=x-y,\qquad u=y-z,\qquad v=z+x
\ee
are useful. The $L_{-1}$ condition then becomes
\be
\Big(\frac{\p}{\p x} + \frac{\p}{\p y} + \frac{\p}{\p z}\Big) \mbold{f} =
2\frac{\p}{\p v}\mbold{f} = 0.
\ee
Hence $\mbold{f}=\mbold{f}(t,u,J,K,L)$. The $L_0$ and $L_1$ conditions then read
\be\label{eqn2}
\Big( h_1+h_2+h_3+J+K+L+t\frac{\p}{\p t}+u\frac{\p}{\p u}\Big) \mbold{f}=0\\ \label{eqn3}
\Big( (v+u+t)(h_1+J) + (v+u-t)(h_2+K) + (v-u-t)(h_3+L) \\+
(u+v)t\frac{\p}{\p t} + (v-t)u\frac{\p}{\p u}\Big) \mbold{f} = 0\nonumber
\ee
respectively. Now, instead of (\ref{eqn2}) and (\ref{eqn3}), one could consider
(\ref{eqn4})$=(t-v)$(\ref{eqn2})$+$(\ref{eqn3}) and
(\ref{eqn5})$=-(u+v)$(\ref{eqn2})$+$(\ref{eqn3}). Since the
transformation is invertible, the conditions (\ref{eqn4}) and
(\ref{eqn5}) are equivalent to the conditions (\ref{eqn2}) and
(\ref{eqn3}). Hence, one has
\be\label{eqn4}
\Big( (2t+u)(h_1+J)+u(h_2+K)-u(h_3+L)+(u+t)t\frac{\p}{\p t}\Big) \mbold{f} = 0\\
\label{eqn5}
\Big( t(h_1+J)-t(h_2+K)-(2u+t)(h_3+L)-(u+t)u\frac{\p}{\p u}\Big) \mbold{f} = 0.
\ee
On expanding $\mbold{f}$ in $J,K,L$, (\ref{eqn4}) gives rise to
$M\times N\times P$ coupled first order differential equations in the variable
$t$. Similarly, (\ref{eqn5}) gives rise to $M\times N\times P$ coupled first order
differential equations in the variable $u$. Each
of (\ref{eqn4}),(\ref{eqn5}) then has $M\times N\times P$ linearly independent
solutions.

Consider the function
\be\label{3ptfn}
\mbold{g}(t,u,J,K,L) &=&
\mbold{C}(J,K,L)t^{-h_1-h_2+h_3-J-K+L}u^{-h_2-h_3+h_1+J-K-L}\times\nonumber \\
&&\qquad (u+t)^{-h_1-h_3+h_2-J+K-L}
\ee
On expanding $\mbold{C}(J,K,L)$ in $J,K,L$, one can see that
$\mbold{C}$ has $M\times N\times P$
components. By direct substitution, $\mbold{g}$ satisfies each of (\ref{eqn4}) and
(\ref{eqn5}). Since $\mbold{g}$ has $M\times N\times P$ free components, one can conclude
that it is the most general expression for the solution of
(\ref{eqn4}) and (\ref{eqn5}). Note that there are no conditions on
$\mbold{C}$, nor are there any conditions on $M,N,P$.

These results appear to be in agreement with the literature, e.g. after
restricting (\ref{3ptfn}) to the rank two case, and the case of primaries not being
pre-logarithmic, the results here match the results of \cite{moghi}.

\section{Four point function}

First consider the `usual' case without Jordan blocks. One wishes to calculate
\be
G^{(4)}(z_1,z_2,z_3,z_4) = \langle \phi_1(z_1)\phi_2(z_2)\phi_3(z_3)\phi_4(z_4)\rangle
\ee
where the $\phi_i$ are quasiprimary fields. The $J_i$ are nilpotent,
although they need not be nilpotent of the same degree, i.e. they
satisfy $J_i^{r_i}=0$, $J_i^{r_i-1}\neq 0$ where the $r_i$ are need
not be the same. In order to perform the calculation, consider the
change of co-ordinates
\be
u=(z_1-z_2),\qquad v=(z_2-z_3),\qquad x =
\frac{(z_1-z_2)(z_3-z_4)}{(z_1-z_3)(z_2-z_4)}, \qquad t = (z_1+z_4).
\ee
The Ward identity for $L_{-1}$ then reads
\be\label{4ptL-1}
\sum_{i=1}^4\frac{\p}{\p z_i}G^{(4)} = 2\frac{\p}{\p t}G^{(4)}=0
\ee
and hence $G^{(4)}=G^{(4)}(u,v,x)$. Defining
$H=\frac{1}{3}\sum_{i=1}^4h_i$, the Ward identity for $L_0$ reads, after using (\ref{4ptL-1}),
\be\label{4ptL0}
\left(\sum_{i=1}^4 h_i+z_i\frac{\p}{\p z_i}\right) G^{(4)} = \left( 3H +
u\frac{\p}{\p u} + v\frac{\p}{\p v}\right) G^{(4)}=0.
\ee
The $L_1$ Ward identity, after use of (\ref{4ptL-1}) and
(\ref{4ptL0}) reads
\be\label{4ptL1}
\sum_{i=1}^4\left( 2h_iz_i + z_i^2\frac{\p}{\p z_i}\right) G^{(4)} =
\Bigg\lbrack (h_1+h_2+h_3-h_4)\left(\frac{uv}{u-x(u+v)}-v\right) + \nonumber\\ (h_1-h_2-h_3-h_4)u
+ (h_1+h_2-h_3-h_4)v + \nonumber\\
\left(\frac{uv}{u-x(u+v)}-v\right) \left(u\frac{\p}{\p u}+v\frac{\p}{\p v}\right) +
uv\left(\frac{\p}{\p u}-\frac{\p}{\p v}\right) \Bigg\rbrack G^{(4)}=0.
\ee
Defining
\be\label{4pt}
F(x,u,v)=G^{(4)}u^{h_1+h_2-H}v^{h_2+h_3-H}(u+v)^{h_1+h_3-H}\left(\frac{uv}{u-x(u+v)}-v\right)^{h_3+h_4-H}\times\nonumber\\
\left(\frac{uv}{u-x(u+v)}\right)^{h_2+h_4-H}\left(\frac{uv}{u-x(u+v)}+u\right)^{h_1+h_4-H}
\ee
one finds that (\ref{4ptL0}) reduces to
\be\label{4ptL0a}
\frac{G^{(4)}}{F}\left(u\frac{\p}{\p u} + v\frac{\p}{\p v}\right) F=0
\ee
and, after use of (\ref{4ptL0a}), that (\ref{4ptL1}) reduces to
\be
\frac{G^{(4)}}{F}uv\left(\frac{\p}{\p u}-\frac{\p}{\p v}\right) F=0.
\ee
Thus, when the points $z_i$ are not coincident, one has
\be
\frac{\p}{\p u}F = 0 = \frac{\p}{\p v}F
\ee
and hence $F=F(x)$. After reorganizing the factors, the general four
point function is then given by (\ref{4pt}). In the logarithmic case,
one wishes to consider the four point function
\be
\mbold{G}^{(4)}(z_1,z_2,z_3,z_4,J_1,J_2,J_3,J_4) = \langle
\phi_1(z_1,J_1)\phi_2(z_2,J_2)\phi_3(z_3,J_3)\phi_4(z_4,J_4)\rangle
\ee
where the $\phi_i$ are logarithmic quasiprimary. Similar to the usual case,
$\frac{\p}{\p t}\mbold{G}^{(4)}=0$. One must now define
$H=\frac{1}{3}\sum_{i=1}^4h_i+J_i$, define
\be
\mbold{F}(x,u,v,
J_i)=\mbold{G}^{(4)}u^{h_1+h_2+J_1+J_2-H}v^{h_2+h_3+J_2+J_3-H}(u+v)^{h_1+h_3+J_1+J_3-H}\times\nonumber\\
\left(\frac{uv}{u-x(u+v)}-v\right)^{h_3+h_4+J_3+J_4-H}
\left(\frac{uv}{u-x(u+v)}\right)^{h_2+h_4+J_2+J_4-H}\times\nonumber\\ \left(\frac{uv}{u-x(u+v)}+u\right)^{h_1+h_4+J_1+J_4-H}
\ee
and use the Ward identities
\be
\left(\sum_{i=1}^4 h_i+J_i+z_i\frac{\p}{\p z_i}\right) \mbold{G}^{(4)} = 0
\nonumber\\ \sum_{i=1}^4\left( 2(h_i+J_i)z_i + z_i^2\frac{\p}{\p z_i}\right)
\mbold{G}^{(4)} = 0
\ee
in a similar manner to the usual case to deduce that
$\mbold{F}=\mbold{F}(x,J_i)$. Thus one finds that\small
\be
\mbold{G}^{(4)}(u,v,x,J_i) = &&
\mbold{F}(x,J_i)u^{-h_1-h_2-J_1-J_2+H}v^{-h_2-h_3-J_2-J_3+H}\times\nonumber\\
&& (u+v)^{-h_1-h_3-J_1-J_3+H}
\left(\frac{uv}{u-x(u+v)}-v\right)^{-h_3-h_4-J_3-J_4+H}\times\nonumber\\ &&
\left(\frac{uv}{u-x(u+v)}\right)^{-h_2-h_4-J_2-J_4+H}\left(\frac{uv}{u-x(u+v)}+u\right)^{-h_1-h_4-J_1-J_4+H}
\ee
which reads in the original co-ordinates (where $x$ is the cross-ratio)
\be\label{4ptfn}
\mbold{G}^{(4)}(z_i,J_i)=\mbold{F}(x,J_i)\prod_{i<k}(z_i-z_k)^{-h_i-h_k-J_i-J_k
+ H}
\ee\normalsize
is the most general logarithmic four point function permitted by
M\"obius symmetry. Note that since the Jordan blocks satisfy $J_i^{r_i}=0$, $J_i^{r_i-1}\neq
0$, then $\mbold{F}$ represents $r_1r_2r_3r_4$ functions of
cross-ratios $x$. On expanding into components, $\mbold{G}^{(4)}$ will
contain logarithms that mix the components of $\mbold{F}$ amongst the
various individual four point functions.

It is instructive to compare this result to examples in the
literature\cite{flohr1}, where the actual primary fields in a Jordan
block are not pre-logarithmic. $\mbold{F}$ can be expanded as $\mbold{F}(x) = F_0(x) +
\sum_{i=1}^4 J_iF_i(x)\ldots$. Taking $F_0=0$ and $F_1=F_2$, one finds
that
\be
G_{12} &=& \frac{1}{3}\left(\prod_{i<j}z_{ij}^{\mu_{ij}}\right) \Big(3F_{12}
+ F_1(-2l_{12} + l_{13} + l_{14} -2l_{23}-2l_{24}+l_{34})\nonumber\\
&& +
F_2(-2l_{12}-2l_{13}-2l_{14}+l_{23}+l_{24}+l_{34})\Big)\nonumber\\ &=&
\frac{1}{3}\left(\prod_{i<j}z_{ij}^{\mu_{ij}}\right) \left(3F_{12} +
F_1\left( -6l_{12} + \log (x) + \log\left(\frac{x}{1-x}\right)\right)\right)
\ee
where $z_{ij}=z_i-z_j$, $\mu_{ij} = -h_i-h_j+\frac{1}{3}\sum_{k=1}^4 h_k$ and $l_{ij} = \log
(z_i-z_j)$. Thus, one finds that logarithms of the cross ratio can appear.

\section{$c=0$ disordered systems}

One starts with a Virasoro OPE, with vanishing central charge
\be\label{TT}
T(z)T(w) = \frac{2T(w)}{(z-w)^2} + \frac{\p T(w)}{(z-w)} + \ldots .
\ee
So as not to let $L_{-2}\ket{0} = \ket{T}$ decouple, leaving a trivial
theory, one can try to realize the theory with a logarithmic partner field
\be\label{Tt}
T(z)t(w) = \frac{b}{(z-w)^4} + \frac{2t(w)+T(w)}{(z-w)^2} +
\frac{\p t(w)}{(z-w)}+\ldots
\ee
where $b$ is some undefined constant. Note, that since $\langle
T(z)t(w)\rangle\neq 0$ is required, this implies that
$\braket{0}{0}\neq 0$. From comparing with the two point function (\ref{2ptfn}) of a
logarithmic quasiprimary field, this implies that the identity
operator cannot be a part of a Jordan block. Hence, there are three
fundamental fields in the theory; the identity $1$, the Virasoro
generator $T$, and the Virasoro generator's logarithmic partner $t$. For this system (\ref{TT}),
(\ref{Tt}) and the field content of $\{ 1,T,t\}$ will be the only
facts assumed. One can then try and construct an operator algebra,
and ask if that algebra is consistent. One can immediately read of the two point functions
\be
\langle T(z)t(w) \rangle = \frac{b}{(z-w)^4},\qquad \langle
T(z)T(w) \rangle = 0
\ee
and notice that $\ket{T}$ has a non-trivial inner product with
$\ket{t}$, and hence cannot decouple. One can then use global conformal symmetry transformations on $\langle
t(z)t(w)\rangle$, which can be obtained from (\ref{Tt}), to deduce
\be\label{tt2pt}
\langle t(z)t(w)\rangle = \frac{e-2b\log (z-w)}{(z-w)^4}
\ee
where $e$ is an arbitrary constant. Since the generic form of
the two and three point functions are known from global conformal symmetry, one might ask
about the operator content of the theory, in a similar manner that one
does for ordinary CFT. Requiring the fields
to form a closed, associative, commutative operator algebra usually imposes
constraints. From (\ref{tt2pt}), it can be seen that what are normally
structure constants will now become functions. A similar statement
holds for the three point functions. One can denote the structure functions, $C$, as
\be\label{funcOPE}
\phi_i(x) \phi_j(y) = \frac{C_{ij}^{\ \ k}(x,y)\phi_k(y)}{(x-y)^{h_i+h_j-h_k}}+\ldots
\ee
where, as usual, $\ldots$ represent terms with poles in $(x-y)$ of order less
than $h_i+h_j-h_k$. Similarly, the `coefficient' in front of a three point
function can be denoted $C_{ijk}(x,y,z)$. By Taylor expanding (\ref{funcOPE}), one has
\be
\phi_i(x) \phi_j(y) = \frac{C_{ij}^{\ \
    k}(x,y)\phi_k(x)}{(x-y)^{h_i+h_j-h_k}}+\ldots .
\ee
Thus, demanding commutativity of the operator algebra, requires
\be\label{commcndn}
C_{ij}^{\ \ k}(x,y) = C_{ji}^{\ \ k}(y,x) .
\ee
Already, from (\ref{tt2pt}), this can be seen to be too strong a
constraint to impose on all of the structure functions. However, some
of the structure functions do exhibit commutativity, in particular those
that are constant. Labelling the fields $\{ 1,
T, t\}$ as $\{\phi_1, \phi_2, \phi_3\}$, one can see from the OPEs (\ref{TT}), (\ref{Tt}), that
\be\label{structconst}
C_{22}^{\ \ 1}=0,\ C_{22}^{\ \ 2}=2,\ C_{22}^{\ \ 3}=0,\ C_{23}^{\ \
  1}=b,\ C_{23}^{\ \ 2}=1,\ C_{23}^{\ \ 3}=2,\ C_{1j}^{\ \ k}=\delta_j^k
\ee
where the $C_{ij}^{\ \ k}$ are symmetric in $i,j$. From the two point
functions (\ref{tt2pt}), one has
\be
C_{33}^{\ \ 1}(x,y) = e - 2b\log (x-y)
\ee
which represents a structure function not obeying
(\ref{commcndn}). Indeed, the product $t(z)t(w)$ is the only offending
product against commutativity.

Now, using (\ref{TT}) and (\ref{Tt}), one can see that $T(w)+Jt(w) =: T(J,w)$ is a
quasiprimary logarithmic field, and hence its three point function is
given by (\ref{3ptfn}). From (\ref{Tt}) and using the
two point functions, one can deduce that
\be
\lim_{|z-w|\rightarrow 0}\langle T(z)t(w)t(u)\rangle =
\lim_{|z-w|\rightarrow 0}\frac{C_{233}(z,w,u)}{(z-w)^2(w-u)^2(z-u)^2}\nonumber\\
= \lim_{|z-w|\rightarrow 0}\frac{b + 2e - 4b\log (w-u)}{(z-w)^2(w-u)^4} + O((z-w)^{-1}).
\ee
One can do the same with the $\langle T(z)T(w)T(u)\rangle$ and $\langle
T(z)T(w)t(u)\rangle$ correlators. Now, comparing with (\ref{3ptfn}), it can be seen that the
$C_{233}(z,w,u)$, $C_{223}(z,w,u)$ and $C_{222}(z,w,u)$ found are in
fact the most general, even away from $|z-w|\rightarrow 0$. Also,
since $T$ commutes with $t$, this result also yields $C_{323}$ and
$C_{332}$. Similarly, $C_{223} = 2b = C_{232} = C_{322}$,
$C_{222}=0$. Using these numbers, and the general form of the 3 point
function (\ref{3ptfn}), one can deduce that
\be
&&\langle t(x)t(y)t(z)\rangle =
\frac{C_{333}(x,y,z)}{(x-y)^2(y-z)^2(x-z)^2} 
\ee
where
\be
&&C_{333}(x,y,z)= d -
(b+2e)\Big(\log (x-z)+\log (x-y) + \log (y-z)\Big)\nonumber\\&& -2b\Big( \log^2(x-z)
+ \log^2(x-y) + \log^2(y-z) - 2\log(x-y)\log(x-z) \nonumber\\&& -
2\log(x-y)\log(y-z) -2\log(y-z)\log(x-z)\Big) .
\ee
Now,
\be
t(x)t(y) = \frac{C_{33}^{\ \ 1}(x,y)1}{(x-y)^4} + \frac{C_{33}^{\ \
    2}(x,y)T(y)}{(x-y)^2} + \frac{C_{33}^{\ \ 3}(x,y)t(y)}{(x-y)^2} + \ldots
\ee
where $\ldots$ represents terms with at most a simple pole in
$x-y$. Thus
\be
\langle t(x)t(y)T(z) \rangle = \frac{C_{33}^{\ \ 3}(x,y)\langle
  t(y)T(z)\rangle }{(x-y)^2}+\ldots .
\ee
Considering the limit $|x-y|\rightarrow 0$, and using $b\neq 0$, one can deduce that
\be
&C_{33}^{\ \ 3}(x,y) = 1 + 2\frac{e}{b} - 4 \log (x-y).
\ee
Similarly,
\be
\langle t(x)t(y)t(z) \rangle = \frac{C_{33}^{\ \ 2}(x,y)\langle
  T(y)t(z)\rangle }{(x-y)^2} + \frac{C_{33}^{\ \ 3}(x,y)\langle
  t(y)t(z)\rangle }{(x-y)^2}+\ldots
\ee
and the limit $|x-y|\rightarrow 0$ yields
\be
&C_{33}^{\ \ 2}(x,y) = \frac{1}{b}\Big( d - e\left( 1+\frac{2e}{b}\right) -
(b-2e)\log(x-y) - 2b\log^2(x-y)\Big)
\ee
and hence all of the structure constants are obtained. 

Given the field content of the theory, the most general singular terms
that can appear in the $t(x)t(y)$ OPE are given by
\be
t(x)t(y) &=& \frac{C_{33}^{\ \ 1}(x,y)1}{(x-y)^4} + \frac{C_{33}^{\ \
    2}(x,y)T(y)}{(x-y)^2} + \frac{C_{33}^{\ \ 3}(x,y)t(y)}{(x-y)^2}
+\nonumber\\ &&
\frac{A(x,y)\p t(y)}{(x-y)} + \frac{B(x,y)\p T(y)}{(x-y)} + \ldots .
\ee
One can use conformal `invariance' of the theory to obtain $A$ and $B$, i.e. note
\be
\co{L_1}{t(x)t(y)}\ket{0} = \co{L_1}{t(x)}t(y)\ket{0} + t(x)\co{L_1}{t(y)}\ket{0}
\ee
and one can take the OPE and act with $L_1$, or act with $L_1$ then
take the OPE. Comparing the $T$ and $t$ coefficients from these two
calculations yields
\be
&A= \frac{1}{2}\left( 1+\frac{2e}{b} - 4\log (x-y)\right) =
\frac{1}{2}C_{33}^{\ \ 3}\\
&B= \frac{1}{2b}\left( d -
e(1+\frac{2e}{b})-(b-2e)\log(x-y)-2b\log^2(x-y)\right) =
\frac{1}{2}C_{33}^{\ \ 2}
\ee
and thus all singular components of the OPE are known.

These calculations give the operator algebra found in
\cite{kogannich2}, where the algebra was found by different
methods. In particular, since in this note only (\ref{TT}) and
(\ref{Tt}) were needed to obtain the operator algebra, the above calculations answer
an important question - given a $c=0$ system with non-degenerate
vacuum, where the Virasoro
generator $T$ has a logarithmic partner, does one always arrive at the
same operator algebra? Up to parameters $b,d$ ($e$ can be removed by
redefinition of $t$ - see next section), the answer is yes. Note that
$b\neq 0$ has been used, which is necessary for $T$ not to decouple,
as was the original motivation. If one wishes to consider $b=0$, and
$T$ not decoupling, then from looking at the two point function
(\ref{gen2ptfn}), larger rank Jordan blocks will be needed.

\section{Associativity}

There are a number of free parameters, namely $b,e,d$. For $b\neq 0$, $t$ can be
redefined by $t\mapsto t - \frac{e}{2b}T$ which leaves (\ref{Tt})
unchanged. However, it does set $e=0$ in (\ref{tt2pt}). Whilst
strictly, one needs to look at the four-point function to understand
the constraints arising from associativity, by looking at an analogous
algebraic structure, one can formally solve for the
constraints. Consider an associative algebra spanned by a finite number of
`fields' $\{ A_I\}$, over the polynomial ring $\mathbb{C}\lbrack
x\rbrack$. Multiplication is given by
\be
A_IA_J = \sum_P C_{IJ}^{\ \ P}(x)A_P.
\ee
This mimics the operator algebra structure, with the logarithms given
by $x$. Associativity of this algebra imposes
\be
\sum_P C_{IJ}^{\ \ P}(x)C_{PK}^{\ \ L}(x) = \sum_P C_{IP}^{\ \
  L}(x)C_{JK}^{\ \ P}(x)
\ee
on the structure constants. One can now try to impose this structure
on the algebra at hand
\be\label{assoccndn}
\sum_p C_{ij}^{\ \ p}(x)C_{pk}^{\ \ l}(x) = \sum_p C_{ip}^{\ \
  l}(x)C_{jk}^{\ \ p}(x).
\ee
Requiring associativity,
and setting $(i,j,k,l)=(2,2,3,2)$ in (\ref{assoccndn}) and using only
(\ref{structconst}) which were assumed at the beginning of the
calculation, yields $b+2=0$. One can check that for $b+2=0$, the
identity holds for all $(i,j,k,l)$. 

One can ask if this result can be
reproduced using the four point function. Since $T$ has its only
non-zero two point function when the other field in the correlator is
$t$, $(i,j,k,l)=(2,2,3,2)$ is
equivalent to studying the four point function $\langle
T(x)T(y)t(w)t(z)\rangle$. Now, $T(x)$ has a mode expansion
$T(x)=\sum_{m\in\mathbb{Z}}L_mz^{-m-2}$. Given this mode expansion,
(\ref{TT}) and (\ref{Tt}) then yield
\be\label{commrelns}
&&\co{L_m}{T(z)} = 2(m+1)z^mT(z) + z^{m+1}\p T(z)\\ \label{commrelns2}
&&\co{L_m}{t(z)} = \frac{b}{6}m(m^2-1)z^{m-2} + (m+1)z^m(2t(z)+T(z)) +
z^{m+1}\p t(z).
\ee
Using these commutation relations and
\be\label{vacprops}
\bra{0}L_m=0 \textrm{ for } m\leq 1,\qquad L_n\ket{0}=0\textrm{ for }
n\geq -1, 
\ee
one can calculate $\langle T(x)T(y)t(w)t(z)\rangle$ for $|x|>|y|$ in
terms of two and three point functions, and analytically continue in $(x-y)$
to obtain the full four point function. This yields
\be\label{TTtt}
\langle T(x)T(y)t(w)t(z)\rangle &=& \frac{2\langle
  T(y)t(w)t(z)\rangle}{(x-y)^2} + \frac{\langle\p T(y)t(w)t(z)\rangle}{(x-y)} +
\frac{b\langle T(y)t(z)\rangle}{(x-w)^4} + \nonumber\\
&&\frac{2\langle T(y)t(w)t(z)\rangle}{(x-w)^2} + \frac{\langle
  T(y)T(w)t(z)\rangle}{(x-w)^2} + \frac{\langle T(y)\p
  t(w)t(z)\rangle}{(x-w)} +\nonumber \\ &&\frac{b\langle
  T(y)t(w)\rangle}{(x-z)^4} + \frac{2\langle
  T(y)t(w)t(z)\rangle}{(x-z)^2} + \frac{\langle
  T(y)t(w)T(z)\rangle}{(x-z)^2} + \nonumber\\ &&\frac{\langle T(y)t(w)\p
  t(z)\rangle}{(x-z)}.
\ee
The identity in question comes down to taking the OPE of $T(x)$ with
$T(y)$ and then evaluating the four point function, and comparing this
to taking the $T(y)t(w)$ OPE and evaluating the four point
function. To this end, consider
\be\label{c22c33}
\lim_{|x-y|\rightarrow 0}\langle T(x)T(y)t(w)t(z)\rangle = \lim_{|x-y|\rightarrow 0}\langle
\left( \frac{2T(y)}{(x-y)^2} + \frac{\p T(y)}{(x-y)}
\right)t(w)t(z)\rangle + O((x-y)^0)
\ee
which, in the limit $|x-y|\rightarrow 0$, agrees with
(\ref{TTtt}). Similarly, consider\small
\be\label{c23c23}
\lim_{|y-w|\rightarrow 0}\langle T(x)T(y)t(w)t(z)\rangle &=& \lim_{|y-w|\rightarrow 0}\langle
T(x)\left( \frac{b}{(y-w)^4} + \frac{2t(w)+T(w)}{(y-w)^2} + \frac{\p
  t(w)}{(y-w)}\right)t(z)\rangle \nonumber \\&& + O((y-w)^0).
\ee\normalsize
This does not obviously agree with (\ref{TTtt}). However, taking
$|y-w|=|x-z|=\epsilon$ and using the expressions for the three point
functions (or, in the limit using, the $T(x)t(z)$ OPE and the two point
functions), one can show that as $\epsilon\rightarrow 0$,
both (\ref{c23c23}) and (\ref{TTtt}) yield
\be
\lim_{\epsilon\rightarrow 0} \frac{b^2}{(y-w)^4(x-z)^4} +
\frac{2(b+2e-4b\log(w-z)) + 2b}{(y-w)^2(x-z)^2(x-w)^2(w-z)^2} + O(\epsilon^{-3})
\ee
and hence, in the limit, both functions agree. In particular, there is
no restriction on $b$. Hence, the notion of $b+2=0$ is really an
illusion from performing too na{\"i}ve a calculation, and missing out
the conformal blocks in (\ref{assoccndn}). However, since na{\"i}vely the only thing
stopping associativity seemed to be $b+2\neq 0$, one might suspect that the
algebra is in fact associative for arbitrary $b$. Of course, to check this properly
would require checking all of the four point functions, but since
$t(z)$ does not appear to have a mode expansion, it is not obvious to
the author how to compute $\langle t(x)t(y)t(w)t(z)\rangle$. However,
using the mode expansion of $T$ and the three point functions, all the
other four point functions can be calculated, and
can used to check associativity. Unfortunately, as can be seen from the general form of the
four point function obtained from only M{\"o}bius symmetry (\ref{4ptfn}), the calculable
four point functions are not quite enough to obtain the $\langle
tttt\rangle$ correlation function - there is still one arbitrary
function of the cross-ratio that needs to be found. As such, a full
calculation to check the associativity of the operator algebra is
still an open problem. Nonetheless, one can check the calculable four
point functions to see if any yield non-associativity (as done in the
appendix), and they do not. In particular, they do not give any
constraints on $b$.

Thus, assuming just (\ref{TT}), (\ref{Tt}), one can deduce a general
operator algebra, which one might expect to be associative, which has two free components, namely $b,d$.

\section{Conclusions}

Purely from the Ward identities for M{\"o}bius symmetry, the general
three and four point functions were obtained. Whilst the author has
not done it, one should be able to find the higher $N$ point functions
by a similar calculation, with the change of co-ordinates involving
more cross-ratios. If the primary fields in
the Jordan blocks of the logarithmic primaries are not
pre-logarithmic, then further constraints appear on the three and four
point functions, which amounts to setting some constants to zero in
the three point case, and some functions of cross-ratios to zero in
the four point case.

Taking the example of $c=0$ systems with non-degenerate vacua, the
three point function proved to be extremely useful in finding the
operator algebra, and the four point function was useful in the
analysis of associativity.

In the analysis of $c=0$ systems, it was found that just assuming that
$T$ had a logarithmic partner and that $T$ did not decouple, one could deduce
that the identity was not a member of a Jordan block, and deduce the full
operator algebra, which was parameterized by two
constants, $b$ and $d$. This result matches that of \cite{kogannich2},
although the derivation here is different, and possibly more general. On a formal level, the associativity
conditions were checked. On a more precise level, almost all of the associativity conditions were
checked. Since the author was unable to obtain the $\langle
tttt\rangle$ correlator, it still remains an open question as to
whether or not this four point function yields any conditions on
associativity. However, all the other four point functions
could be found, and were tested to see if they gave signs of
non-associativity. They did not. These findings suggest that $c=0$
systems with non-degenerate vacua may well give consistent field theories, although the
final steps of the argument remain unfinished.

The operator algebra obtained differed quite significantly to those in
normal CFT, in that due to logarithms in the OPE, it is not obvious
how to relate $t(z)t(w)$ to $t(w)t(z)$. One resolution might be to
define $\bar{\p}t(z,\bar{z})$ as an antiholomorphic weight $1$ primary
field, similar to a free boson, as touched on in \cite{kogannich1}. The
logarithms would then appear as $\log|z-w|$ rather than $\log(z-w)$,
and hence $t(z)t(w)$ might be symmetric. The logical end to this
input, and the resultant operator algebra, is not something the author
has done.

The author hopes that this note has given a good illustration of how
BPZ \cite{bpz} machinery can be generalized to the logarithmic scenario.

\subsection*{Acknowledgements}
The author would like to thank Michael Flohr, for getting him
interested in $c=0$ systems, and for numerous helpful discussions
during this work. The author would also like to thank James Lucietti,
for comments on the manuscript.

\subsection*{Note added}
During the writing of this manuscript, \cite{ras1} was released,
where, in the case of rank two Jordan blocks and without assuming
anything about the operator algebra, the three point function
was obtained. After restricting the three point function (\ref{3ptfn})
to the rank two case, (\ref{3ptfn}) then agrees with the three point
function found in \cite{ras1}.

\appendix

\section{c=0 four point functions and associativity}

Since if there is more than one identity operator in correlator, the
calculation will clearly give an answer of associativity, only the
case of $\leq 1$ operator in the correlator being the identity will be
considered. Since $C_{22}^{\ \ 3}=C_{22}^{\ \ 1}=0$, and the
correlators $\langle TT\rangle$, $\langle
TTT\rangle$ and $\langle TTTT\rangle$ are zero, the correlators $\langle
TTT1\rangle$ and $\langle TTTT\rangle$ will yield
associativity. Checking that the $\langle ttt1\rangle$, $\langle
Ttt1\rangle$ and $\langle TTt1\rangle$ correlators yield associativity are not
difficult or long calculations, and indeed they do.

Using (\ref{commrelns}), (\ref{commrelns2}), (\ref{vacprops}), as
before, one can compute $\langle T(x)T(y)T(z)t(w)\rangle$ to give
(after noting the $\langle TT\rangle$ and $\langle TTT\rangle$
correlators are zero)
\be\label{TTTt}
\langle T(x)T(y)T(z)t(w)\rangle &=& \frac{2\langle
  T(y)T(z)t(w)\rangle}{(x-y)^2}  + \frac{\langle
  \p T(y)T(z)t(w)\rangle}{(x-y)} + \frac{2\langle
  T(y)T(z)t(w)\rangle}{(x-z)^2}  + \nonumber\\ &&\frac{\langle
  T(y)\p T(z)t(w)\rangle}{(x-z)} + \frac{2\langle
  T(y)T(z)t(w)\rangle}{(x-w)^2}  + \frac{\langle
  T(y)T(z)\p t(w)\rangle}{(x-w)} 
\ee
which clearly agrees with the $|x-y|\rightarrow 0$ limit and the
$T(x)T(y)$ OPE
\be
\lim_{|x-y|\rightarrow 0}\langle T(x)T(y)T(z)t(w)\rangle &=&
\lim_{|x-y|\rightarrow 0} \langle\left( \frac{2T(y)}{(x-y)^2} + \frac{\p T(y)}{(x-y)}
\right)T(w)t(z)\rangle + \nonumber\\ && O((x-y)^0).
\ee
Considering $|y-z|=|x-w|=\epsilon$ and $\epsilon\rightarrow 0$, both
(\ref{TTTt}) and using the $T(y)T(z)$ OPE with three point functions,
i.e.,
\be
\lim_{|y-z|\rightarrow 0}\langle T(x)T(y)T(z)t(w)\rangle &=& \lim_{|y-z|\rightarrow 0}\langle
T(x)\left( \frac{2T(z)}{(y-z)^2} + \frac{\p
  T(z)}{(y-z)}\right)t(w)\rangle\nonumber\\ &&+ O((y-z)^0)
\ee
yield
\be
\lim_{\epsilon\rightarrow 0} \langle T(x)T(y)T(z)t(w)\rangle =
\lim_{\epsilon\rightarrow 0} \frac{4b}{(x-w)^2(y-z)^2(z-w)^2(x-z)^2} + O(\epsilon^{-3})
\ee
and hence agree in this limit. 

Using the same techniques, the $\langle
Tttt\rangle$ correlator is given by\small
\be
\langle T(x)t(y)t(w)t(z)\rangle &=& \frac{b\langle
  t(w)t(z)\rangle}{(x-y)^4} + \frac{2\langle t(y)t(w)t(z)\rangle +
  \langle T(y)t(w)t(z)\rangle}{(x-y)^2} + \frac{\langle\p
  t(y)t(w)t(z)\rangle}{(x-y)} +\nonumber\\
  && \frac{b\langle
  t(y)t(z)\rangle}{(x-w)^4} + \frac{2\langle t(y)t(w)t(z)\rangle +
  \langle t(y)T(w)t(z)\rangle}{(x-w)^2} + \frac{\langle
  t(y)\p t(w)t(z)\rangle}{(x-w)} +\nonumber\\
  && \frac{b\langle
  t(y)t(w)\rangle}{(x-z)^4} + \frac{2\langle t(y)t(w)t(z)\rangle +
  \langle t(y)t(w)T(z)\rangle}{(x-z)^2} + \frac{\langle
  t(y)t(w)\p t(z)\rangle}{(x-z)}
\ee\normalsize
which in the limit $|x-y|\rightarrow 0$ agrees with the OPE
\be
\lim_{|x-y|\rightarrow 0}\langle T(x)t(y)t(w)t(z)\rangle
&=&\lim_{|x-y|\rightarrow 0} \langle\left( \frac{b}{(x-y)^4} +
\frac{2t(y)+T(y)}{(x-y)^2} + \frac{\p t(y)}{(x-y)}\right)
t(w)t(z)\rangle \nonumber\\ &&+ O((x-y)^0).
\ee
To analyze the other OPE, one must once again set $|y-w| = |x-z| =
\epsilon$ and take the limit and OPE. In this case, once again the
four point function and OPE calculation agree, with leading order
behaviour
\be
\lim_{\epsilon\rightarrow 0} \frac{b(e-2\log(y-w))}{(y-w)^4(x-z)^4} +
\frac{b+2e+2d -(4b+8e)\log(y-x)-
  (4e+6b)\log(y-w)}{(y-w)^2(x-z)^2(x-y)^4} \nonumber\\
+\frac{16b\log(y-x)\log(y-w)-4b\log^2(y-w)}{(y-w)^2(x-z)^2(x-y)^4} + o(\epsilon^{-3-\frac{1}{2}})
\ee
where the $\epsilon^{-\frac{1}{2}}$ is to suppress the $\log$s in the
limit.

\end{document}